\documentclass{appolb}
\usepackage{graphicx}
\usepackage{amssymb}
\usepackage{amsmath}
\usepackage{bm}
\newcommand{\sgn}{\text{sgn}}
\newcommand{\arccot}{\text{arccot}}

\begin{document}
\title{Incompleteness in mathematical physics.%
\thanks{submitted 24-4-2019}%
}
\author{Han Geurdes
\address{Geurdes data science KvK64522202 CvdLinstraat 164 2593NN Den Haag Netherlands}
\\
{Koji Nagata
}
\address{Department of Physics, Korea Advanced Institute of Science and Technology, Daejeon, Korea }
\\
{Tadao Nakamura
}
\address{Department of Information and Computer Science,  
Keio University, 
3-14-1 Hiyoshi, 
Kohoku-ku,
Yokohama, 
Japan }
\\
{Ahmed Farouk
}
\address{Computer Sciences Department, 
Faculty of Computers and Information, Mansoura University,
Scientific Research Group, Egypt }
}
\maketitle
\begin{abstract}
In the paper it is demonstrated that Bell's theorem is an unprovable theorem.
\end{abstract}
  
\section{Introduction}
Let us start our paper with a quote from professor Friedman's last lecture \cite{FriedLast}.(cit $\dots$) Most [mathematicians] intuitively feel that the great power and stability of some "rulebook for mathematics" is an important component of
their relationship with mathematics. The general
feeling is that there is nothing substantial to be
gained by revisiting the commonly accepted rule book $\dots \quad$ .

The incompleteness demonstrated in this paper is based on the following. Suppose we have a set of axioms $\mathcal{A}$ and a set of derivation rules $\mathcal{S}$.
The axioms and derivation rules constitute in our case concrete mathematics.
With the derivation rules in syntax $\mathcal{S}$, say with a subset $\mathcal{S}_1 \subset \mathcal{S}$, a theorem $\mathcal{T}$ can be derived from the axioms $\mathcal{A}$ or from $\mathcal{A}_1 \subset \mathcal{A}$.
The axioms and derivation rules are \emph{negation incomplete} when the negation of the theorem $\mathcal{T}$, i.e. $\neg\mathcal{T}$, can be derived as well from the same axioms $\mathcal{A}$, or $\mathcal{A}_2 \subset \mathcal{A}$ using the same syntax $\mathcal{S}$ or a subset $\mathcal{S}_2\subset \mathcal{S}$. 
Please do observe that possibly $\mathcal{S}_1\neq \mathcal{S}_2$ and/or $\mathcal{A}_1\neq \mathcal{A}_2$. 
G{\"o}del demonstrated that negation incompleteness holds true for every abstract mathematical set of axioms $\mathcal{A}$ with derivation rules $\mathcal{S}$ that is a sufficient smooth and formal system \cite{YessVolp}. 
Let us also refer to a translation of G{\"o}dels work with an extensive introduction \cite{MeltBrai}

In the present paper we will demonstrate negation incompleteness for Bells formula. The theorem $\mathcal{T}$ is "CHSH is valid". The second branch of the incompleteness comprises a counter model where $\neg\mathcal{T}$ is "CHSH is invalid". Both branches will be demonstrated true with the system of concrete mathematics having axioms $\mathcal{A}$ and derivation rules $\mathcal{S}$. This comprises concrete mathematical incompleteness.

In 1964, John Bell wrote a paper \cite{2} about the possibility of hidden variables \cite{1} causing the entanglement correlation $E(a,b)$ between two particles. In the present paper, an inconsistency similar to concrete mathematical incompleteness \cite{ProfFriedman}, will be demonstrated from his theorem. The argument for mathematical  incompleteness is to give proof {\it and} refute with known concrete mathematical means the mathematical statement of Bell. The authors are aware of the scepsis this may raise with certain readers. However, scepsis is simply not enough to push our proof of inconsistency aside and do "business as usual" with Bell's formula.

Bell, based his hidden variable description on particle pairs with entangled spin, originally formulated by Bohm \cite{3}. Bell used  hidden variables $\lambda$ that are elements of a universal set $ \Lambda$ and are distributed with a density $\rho(\lambda)\geq 0$. 
Suppose, $E(a,b)$ is the correlation between measurements with distant A and B that have unit-length, i.e. $||a||=||b||=1$, real 3 dim parameter vectors $a$ and $b$.
Although it is mathematics that we are dealing with, it is good to look at the basic physics requirements because they determine the boundaries of application.
The basic physics experiment is as follows: 
Suppose on the A side we have measurement instrument A with parameter vector $a$. On the B-side we have measurement instrument B with parameter vector $b$. There is a (Euclidean) distance $d(A,B)>0$ between instruments $A$ and $B$ which can be large if necessary.
In between the two instruments there is a source $\Sigma$ generating particle pairs. We have, $d(\Sigma,A)=d(\Sigma,B)=\frac{1}{2}d(A,B)$. One particle of the pair is sent to A the other particle of the pair is sent to B. The physics of the two particles of the pair is such that they are entangled, \cite{3},\cite{4}.

Then with the use of the $\lambda$ we can write down the classical probability correlation between the two simultaneously measured particles. This is what we will call Bell's formula.
\begin{equation}\label{1}
E(a,b)=\int_{\lambda \in \Lambda} \rho(\lambda)A(a,\lambda)B(b,\lambda) d\lambda
\end{equation}
Note that if $\ell$ is the short-hand notation for the random variable(s), the $E(a,b)$ simply is the expectation value of the product of two $\{-1,1\}$ functions, 
$A(a,\lambda)$ and $B(b,\lambda)$. It can be written as $E(a,b)=E_{\ell}\left( A(a,\ell)B(b,\ell)\right)$. We are looking at a special case of covariance computation \cite{Hogg}.

In (\ref{1}) we therefore must have  $\int_{\lambda \in \Lambda} \rho(\lambda) d\lambda =1$. The integration $\int_{\lambda \in \Lambda}$ can contain as many as we please, variables and arbitrary space $\Lambda$. The density $\rho \geq 0$ also has a very general form. 
\subsection{Proof along the lines of the CHSH}
From  (\ref{1}) an inequality for four setting combinations, $a,b,c$ and $d$ can be derived as follows
\begin{eqnarray}\label{2}
E(a,b)-E(a,c)=
\int_{\lambda \in \Lambda} d\lambda \rho(\lambda)  A(a,\lambda)B(c,\lambda)A(d,\lambda)B(c,\lambda)
-\nonumber\\
\int_{\lambda \in \Lambda} d\lambda \rho(\lambda)  A(a,\lambda)B(b,\lambda)A(d,\lambda)B(b,\lambda) 
+\nonumber \\ 
\int_{\lambda \in \Lambda}d \lambda \rho(\lambda)A(a,\lambda)B(b,\lambda)
-\nonumber\\
\int_{\lambda \in \Lambda}d \lambda \rho(\lambda)A(a,\lambda)B(c,\lambda)
\end{eqnarray}
because, $\{B(c,\lambda)\}^2=\{B(b,\lambda)\}^2=1$. From this it follows
\begin{eqnarray}\label{3}
E(a,b)-E(a,c)=\int_{\lambda \in \Lambda} d\lambda \rho(\lambda)A(a,\lambda)B(b,\lambda)\left\{ 1- A(d,\lambda)B(b,\lambda)\right\}+\nonumber \\
\int_{\lambda \in \Lambda} d\lambda \rho(\lambda)\left(-A(a,\lambda)B(c,\lambda) \right)\left\{ 1- A(d,\lambda)B(c,\lambda)\right\}
\end{eqnarray}
Hence, because $1-A(x,\lambda)B(y,\lambda) \geq 0$ for all $x,y$ with $||x||=||y||=1$ and $A(a,\lambda)B(b,\lambda) \leq 1$ together with $-A(a,\lambda)B(c,\lambda)\leq 1$, it can be derived that
\begin{equation}\label{4}
E(a,b)-E(a,c)\leq 2 - E(d,b)-E(d,c) 
\end{equation}
Or,
\begin{equation}\label{4a}
S(a,b,c,d)=E(a,b)+E(d,b)+E(d,c)-E(a,c) \leq 2.
\end{equation}
Note, no physics assumptions were employed in the derivation of (\ref{4}). It is pure mathematics.  
Suppose, further, that if we select for $a,b,c$ and $d$
\begin{eqnarray}\label{5}
a=\frac{1}{\sqrt{2}}\left(1,0,1\right),\,d=\left(\frac{1}{2},\frac{1}{\sqrt{2}},-\frac{1}{2}\right)\nonumber \\
b=\left(1,0,0\right),\, c=\left(0,0,-1\right)
\end{eqnarray}
then $E(x,y)$ cannot be the inner product of the two vectors because, $a\cdot b= \frac{1}{\sqrt{2}}, d\cdot b= \frac{1}{2}, d\cdot c= \frac{1}{2} $ and $a\cdot c = -\frac{1}{\sqrt{2}}$.  Hence, 
\[ S(a,b,c,d)=(a\cdot b)+ (d\cdot b) + (d\cdot c) -(a\cdot c) =\frac{1}{\sqrt{2}}+\frac{1}{2}+\frac{1}{2}-\left(-\frac{1}{\sqrt{2}}\right) =1+\sqrt{2}>2 
\]
In \cite{4} Peres gives supporting argumentation to the form, $S(a,b,c,d)\leq 2$  derived here. So we can be sure (\ref{4}) and $S(a,b,c,d)\leq 2$, are a generally valid expression for {\it all} possible models under the umbrella of (\ref{1}).

\section{Counter proof with a specific model}\label{Prel}
In this section we will demonstrate that $E(x,y)$ can arbitrarily close approximate the inner product $x \cdot y$.  As a reminder, both $x\in\mathbb{R}^3$ and $y\in\mathbb{R}^3$ are unit length parameter vectors, hence, $E(x,y)\in[-1,1]$. Although the physical details are unimportant, they can be verified to be within the bounds of applicability of Bell's formula (\ref{1}). 

\subsection{Preliminaries}
The model to be developed here follows the basic physical requirements of a local model. The requirements follow from looking at the physics experiment. In instrument A a set of hidden variables is supposed. Similarly, a set of hidden variables is supposed to reside in instrument B. 
The instruments are, as in the previous section, represented in the formulae by functions $A(x,\lambda_{I},\chi)$ and $B(y,\lambda_{II},\chi)$. 
The (arrays of) hidden variables $\lambda_{I}$ and $\lambda_{II}$ are independent. 
A third set of hidden variables, denoted here by $\chi$, are "carried by the particles". 
The $\chi$ have a Gaussian density. 
The $\chi$ variables are independent of $\lambda_{I}$ and $\lambda_{II}$. 
Moreover,  $\lambda_{I}$ and $\lambda_{II}$ are independent. 
Looking at (\ref{1}) we see that $\lambda=(\lambda_{I},\lambda_{II},\chi)$. 
Hence, looking at (\ref{1}), $A(a,\lambda)=A(a,\lambda_{I},\chi)$, $B(b,\lambda)=B(b,\lambda_{II},\chi)$ and $\rho(\lambda)=\rho(\lambda_{I},\lambda_{II},\chi)$. 
This is a local and physically possible situation. 
Although the proof we deliver here is purely mathematics, the necessary basic physical requirements are fulfilled in the model. 

It must be stressed that, in anticipation of a more detailed definition below, the mathematical form of the probability density $\rho(\lambda_{I},\lambda_{II},\chi)$ remains {\it fixed} all the time. 
This can be easily verified in the section below devoted to the probability density. 

It will be shown that the argument of Bell is based on negation incompleteness. In other words, we will show $S(a,b,c,d)> 2$  from the same formula (\ref{1}) that with the same physical requirements gives, along the branch of Bell's argumentation, $S(a,b,c,d)\leq 2$. 

\subsection{Probability density}\label{ProbDens}
Let us in the first place define a probability density $\rho$ based upon two separate $\lambda$'s and on $\left(\chi_1,\chi_2,\chi_3\right)$. Suppose, $\alpha$ is a variable to indicate the two separate systems of hidden variables. Let us denote them with $I$ and $II$, i.e., $\alpha \in \{I,II\}$. Then,
\begin{equation}\label{6}
\lambda_{\alpha}=\left(x_{\alpha},\mu_{1,\alpha},\mu_{2,\alpha},\mu_{3,\alpha}, \tau_{\alpha},n_{\alpha}\right)\in \mathbb{R}^6
\end{equation}
For $\lambda_{I}$ we define a density $\rho_{I}=\rho_{I}(\lambda_{I})$ and for $\lambda_{II}$ a density $\rho_{II}(\lambda_{II})$. The roman indices refer to the two different wings of the Bell experiment.
\subsubsection{The $\chi$ variables}
For $\vec{\chi}=\left(\chi_1,\chi_2,\chi_3\right)\in \mathbb{R}^3$ let us  define the Normal Gaussian density
\begin{equation}\label{7}
\rho_{Norm}=\rho_{Norm}\left(\chi_1,\chi_2,\chi_3\right)=\left(\frac{1}{2\pi}\right)^{3/2}\exp\left[ - \sum_{k=1}^3 \frac{\chi_k^2}{2}\right]
\end{equation}
The integration of the normal density is, $\int_{-\infty}^{\infty}d\chi_1\int_{-\infty}^{\infty}d\chi_2 \int_{-\infty}^{\infty}d\chi_3$ and is denoted with brackets, $\left\langle \cdot \right\rangle_{Norm}$ such that e.g. $\left\langle \rho_{Norm} \right\rangle_{\mathbb{R}^3}=1$. 
This enables us to formally write the total density as 
\begin{equation}\label{8}
\rho(\lambda_{I},\lambda_{II},\vec{\chi})=\rho_{I}(\lambda_{I})\rho_{II}(\lambda_{II})\rho_{Norm}(\vec{\chi})
\end{equation}
The density defined in (\ref{8}) should fulfill the requirements alluded to in the previous section devoted to the requirements of the physics behind the model. The $\chi$ are mutually independent and are independent of the "instrument variables" $\lambda_{I}$ and $\lambda_{II}$.
Subsequently, let us turn to the use of the $\chi $ variables in the model. 

Let us, firstly,  define the Heaviside function $H(x)=1 \Leftrightarrow x \geq 0$ and $H(x)=0 \Leftrightarrow x<0$. In the second place let us define a sign function from the Heaviside,  $\sgn(x)=2H(x)-1$. 
Because of the symmetry of the Gaussian in (\ref{7}), we have in the angular notation of integration for $i,j=1,2,3$ that 
\begin{eqnarray}\label{8a}
\left\langle \sgn(\chi_{i})\sgn(\chi_{j}) \rho_{Norm}\right\rangle_{\mathbb{R}^3}  =
\nonumber \\
\left( \frac{1}{2\pi} \right)^{3/2}\int_{-\infty}^{\infty}d\chi_1\int_{-\infty}^{\infty}d\chi_2 \int_{-\infty}^{\infty}d\chi_3\, \sgn(\chi_{i})\sgn(\chi_{j}) \exp\left[ -  \sum_{k=1}^3 \frac{\chi_k^2}{2}\right] 
= \nonumber\\
\delta_{i,j}
\end{eqnarray}
with, $\delta_{i,j}=1 \Leftrightarrow i=j$ and $\delta_{i,j}=0 \Leftrightarrow i\neq j$.
\subsubsection{Definition of $ \rho_{\alpha}(\lambda_{\alpha}),\,\,\alpha\in\{I,II\} $}
Here we turn to the densities,  $ \rho_{\alpha}(\lambda_{\alpha}),\,\,\alpha\in\{I,II\} $.
The $\rho_{\alpha}(\lambda_{\alpha})$ is a product of five factors, $\rho_{\alpha}^{r}$, $r=0,1,2,3,4$.
We have, for $T \in \mathbb{N}$ and $T>>16$
\begin{eqnarray}\label{9}
\rho_{\alpha}^{0}=\frac{1}{16T\left(1-\frac{4}{T}\right)}\nonumber\\
\rho_{\alpha}^{1}=\rho_{\alpha}^{1}(x_{\alpha})=H\left(\frac{1}{4}+x_{\alpha}\right)H\left( -\frac{1}{T}-x_{\alpha}\right)+H\left(\frac{1}{4}-x_{\alpha}\right)H\left( -\frac{1}{T}+x_{\alpha}\right)\nonumber\\
\rho_{\alpha}^{2}=\rho_{\alpha}^{2}(\vec{\mu}_{\alpha})=\prod_{k=1}^3 H\left( 1+\mu_{k,\alpha}\right)H\left( 1- \mu_{k,\alpha}\right) \nonumber \\
\rho_{\alpha}^{3}=\rho_{\alpha}^{3}(\tau_{\alpha})=H\left( T+\tau_{\alpha}\right)H\left( T- \tau_{\alpha}\right)\nonumber \\
\rho_{\alpha}^{4}=\rho_{\alpha}^{4}(n_{\alpha})=1 \Leftrightarrow n_{\alpha}\in \{0,1\} \,\&\, \rho_{\alpha}^{4}=\rho_{\alpha}^{4}(n_{\alpha})=0 \Leftrightarrow n_{\alpha}\notin \{0,1\}.
\end{eqnarray}
Hence, using (\ref{9}) we then define $\rho_{\alpha}=\prod_{r=0}^4 \rho_{\alpha}^{r}$. 

Subsequently, let us also introduce the angle notation for integration of $\alpha$ densities similar to what we wrote for the Normal density. We have, $T>>16$
\begin{eqnarray}\label{10}
\left \langle  \rho_{\alpha} \dots \right\rangle_{\alpha} =\frac{1}{2^4 T\left(1-\frac{4}{T}\right)}\left( \int_{-\frac{1}{4}}^{-\frac{1}{T}} d x_{\alpha} + \int_{\frac{1}{T}}^{\frac{1}{4}} d x_{\alpha}\right)\prod_{k=1}^3 \int_{-1}^1 d\mu_{k,\alpha} \int_{-T}^{T} d\tau_{\alpha} \sum_{n_{\alpha}=0}^1 \dots
\end{eqnarray}
The previous leads us to $\left\langle \rho_{\alpha}  \right\rangle_{\alpha}=\frac{2}{2^4 T\left(1-\frac{4}{T}\right)}\left( 2^3 \times 2T\right)\left(\frac{1}{2}-\frac{2}{T}\right)=1 $, and, $T \sim $ sufficiently large number. 
Looking at the definition of the total density in (\ref{8}), it can be derived that 
\begin{equation}\label{11}
\int_{\lambda\in \Lambda} d\lambda \rho(\lambda)=\left \langle \left \langle \rho_{I} \right\rangle_{I}\left \langle \rho_{II} \right\rangle_{II}\rho_{Norm}\right\rangle_{\mathbb{R}^3}=\left \langle \rho_{I} \right\rangle_{I}\left \langle \rho_{II} \right\rangle_{II}\left\langle\rho_{Norm}\right\rangle_{\mathbb{R}^3} =1
\end{equation}
Hence, a valid probability density in (\ref{8}) is obtained where use is made of (\ref{7}) and (\ref{9}).
The density given in (\ref{8}) is a valid fixed form density that is completely local.

\subsection{Auxiliary functions}
\subsubsection{ The auxiliary function $\Delta_T(y):$}
Let us in the first place define
\begin{equation}\label{12}
\Delta_T(y)=\frac{2/\pi}{1+T^2y^2}
\end{equation}
Then, because $1+T^2y^2 \geq 1$ for $y\geq 0$, we find that $-T \leq T\Delta_T(y) \leq T$ is valid and so, $\sgn\left( T\Delta_T(y) -\tau_{\alpha}\right)$ can be meaningfully  employed in an integration. 
\begin{eqnarray}\label{13}
\int_{-T}^{T} \sgn\left( T\Delta_T(y) -\tau_{\alpha}\right) d \tau_{\alpha} =
\nonumber\\
\int_{-T}^{T\Delta_T(y)}d\tau_{\alpha}-\int_{T\Delta_T(y)}^{T} d\tau_{\alpha}=
\left(T\Delta_T(y)-(-T)\right)-\left(T-T\Delta_T(y) \right)  =
\nonumber\\ 
2T\Delta_T(y)
\end{eqnarray}
This is true for arbitrary real $y$. Hence, also for $y=x_{\alpha}^2- \frac{1}{T^2}$ the previous is true.
\subsubsection{Elements of the measurement functions:}
In the second place let us define
\begin{eqnarray}\label{14}
\sigma_{a}=\sum_{k=1}^3 a_k \sgn(\chi_k)\nonumber \\
\sigma_{b}=\sum_{k=1}^3 b_k \sgn(\chi_k)
\end{eqnarray}
It is easily demonstrated that $|\sigma_{a}|\leq \sqrt{3}$ and $|\sigma_{b}|\leq \sqrt{3}$.
\subsubsection{Indicators:} 
In the third place, let us define three disjoint partitions of the real interval $[-\sqrt{3},\sqrt{3}]$.
\begin{eqnarray}\label{15}
\begin{array}{ll}
I_1=\{x \in \mathbb{R} \,|\, -\sqrt{3}\leq x < -1\}\\
I_2=\{x \in \mathbb{R} \,|\, -1\leq x \leq 1\}\\
I_3=\{x \in \mathbb{R} \,|\, 1 < x \leq \sqrt{3} \}
\end{array}
\end{eqnarray}
Clearly, $I_1 \cap I_2 =\emptyset$ together with $I_2 \cap I_3 =\emptyset$  and $I_3 \cap I_1 =\emptyset$. With the use of the three disjoint intervals we may employ the following auxiliary function $\iota_k(x)=1 \Leftrightarrow x \in I_k$ and $\iota_k(x)=0 \Leftrightarrow x \notin I_k$. If $\iota_1(x)=1$, then, $\iota_2(x)=\iota_3(x)=0$. If $\iota_2(x)=1$, then, $\iota_1(x)=\iota_3(x)=0$. If $\iota_3(x)=1$, then, $\iota_2(x)=\iota_1(x)=0$.
\subsubsection{Auxiliary functions in measurement functions: }
 Let us fourthly also define functions that will be employed together with the $\iota_k(x)$, with, $k=1,2,3$.
\begin{eqnarray}\label{16}
\begin{array}{ll}
s_{1,\alpha}(z_{\alpha},\lambda_{\alpha})=\left\{n_{\alpha}\sgn\left(z_{\alpha}+1-\mu_{1,\alpha} \right)-\delta_{0,n_{\alpha}} \right\}\sgn \left( T\Delta_T(f_{\alpha}(x_{\alpha}))-\tau_{\alpha} \right)\\
s_{2,\alpha}(z_{\alpha},\lambda_{\alpha})=\sgn\left(z_{\alpha}-\mu_{2,\alpha} \right) \\
s_{3,\alpha}(z_{\alpha},\lambda_{\alpha})=\left\{n_{\alpha}\sgn\left(z_{\alpha}-1-\mu_{3,\alpha} \right)+\delta_{0,n_{\alpha}} \right\}\sgn \left( T\Delta_T(f_{\alpha}(x_{\alpha}))-\tau_{\alpha} \right)
\end{array}
\end{eqnarray}
The $z_{\alpha}$ is a short-hand and follows, $z_{I}=\sigma_a$ and $z_{II}=\sigma_b$, with $\alpha \in \{I,II\}$. 
The $\sigma_a$ and $\sigma_b$ are defined in (\ref{14}). It is quite easily verifiable that $s_{k,\alpha}(z_{\alpha},\lambda_{\alpha}) \in \{-1,1\}$, with $k=1,2,3$. Note, $n_{\alpha} \in \{0,1\}$. In (\ref{16}) the short-hand, $f_{\alpha}(x_{\alpha})\equiv x_{\alpha}^2 -\frac{1}{T^2}$ is employed.

\subsubsection{Measurement functions}
With the use of the previous definitions we are now able to define the measurement functions A and B. 
\begin{eqnarray}\label{17}
A(a,\lambda_{I},\vec{\chi})=\sum_{k=1}^3 \iota_k(\sigma_a)s_{k,I}(\sigma_a,\lambda_{I}),\nonumber \\ B(b,\lambda_{II},\vec{\chi})=\sum_{k=1}^3 \iota_k(\sigma_b)s_{k,II}(\sigma_b,\lambda_{II})
\end{eqnarray}
Because the $\iota_k(x)$, $k=1,2,3$ only have one of them unequal to zero, i.e. the $I_k$ of (\ref{15}) are disjoint, and the $s$ of equation (\ref{16}) are in $\{-1,1\}$, we have that both  $A(a,\lambda_{I},\vec{\chi}) \in \{-1,1\}$ and $B(b,\lambda_{II},\vec{\chi}) \in \{-1,1\}$.
Hence the measurement functions in (\ref{17}) are valid in a Bell correlation $E(a,b)$ such as given in (\ref{1}). 
No deeper physics assumption hides behind this. One simply may select functions that project in $\{-1,1\}$. Bell's formula is general and choice is part of normal concrete mathematics under ZFC. The A and B are called measurement functions but that is totally unimportant to the mathematics to be developed here.

Clearly, we can conclude that our definitions comply to the basic physical requirements of a local model. Hence, the model is allowed in Bell's formula. Note that the measurement representing functions, projecting in $\{-1,1\}$, also follow the basic physical requirements. The derivation of $S(a,b,c,d)\leq 2$ in (\ref{4a}) is therefore possible in this case. We will show that this is just one branch of the argument.
\subsubsection{Evaluation}
Looking at Bell's correlation in (\ref{1}) let us write
\begin{eqnarray}\label{18}
E(a,b)=\left \langle\left \langle \left \langle \rho_{I}  \rho_{II} \rho_{Norm}A(a,\lambda_{I},\vec{\chi}) B(b,\lambda_{II},\vec{\chi})\right\rangle_{I}\right\rangle_{II}\right\rangle_{\mathbb{R}^3} = \nonumber \\
\left\langle \left\langle \rho_{I} A(a,\lambda_{I},\vec{\chi})\right\rangle_{I} \left\langle\rho_{II}B(b,\lambda_{II},\vec{\chi}) \right\rangle_{II}\rho_{Norm} \right\rangle_{\mathbb{R}^3}
\end{eqnarray}
Note, $\lambda_{I}$ is only found in $\rho_{I}$ and $A(a,\lambda_{I},\vec{\chi})$ while  $\lambda_{II}$ is only found in $\rho_{II}$ and $B(b,\lambda_{II},\vec{\chi})$. The $\vec{\chi}$, via the $\sigma_a$ and $\sigma_b$ dependence is shared between functions A and B. Note for completeness that the function $B$ does not depend on $a$ and $A$ does not depend on $b$ which is in accordance with Einstein's locality condition \cite{1}.

In order to have a proper evaluation of the integrals in $\left \langle \rho_{I} A(a,\lambda_{I},\vec{\chi})\right\rangle_{I}$, and $\left \langle \rho_{II}B(b,\lambda_{II},\vec{\chi}) \right\rangle_{II}$ it is sufficient to look at the $A$ side only. 
The $B$ side evaluations obviously follows similar rules.

We can write explicitly for $\left \langle \rho_{I} A(a,\lambda_{I},\vec{\chi})\right\rangle_{I}$
\begin{eqnarray}\label{19}
\left \langle \rho_{I} A(a,\lambda_{I},\vec{\chi})\right\rangle_{I} = 
\nonumber\\
\frac{1}{2^4T\left(1-\frac{4}{T}\right)}\left( \int_{-\frac{1}{4}}^{-\frac{1}{T}} d x_{I} + \int_{\frac{1}{T}}^{\frac{1}{4}} d x_{I}\right)\prod_{k=1}^3 \int_{-1}^1 d\mu_{k,I} \int_{-T}^{T} d\tau_{I} \sum_{n_{I}=0}^1 A(a,\lambda_{I},\vec{\chi})
\end{eqnarray}
As it follows from (\ref{17}), we can have three cases for $\iota_k(\sigma_a)$. 
Suppose, the selection $a$ and the values of $\vec{\chi}$ are such that $\sigma_a$ from equation (\ref{14}) is $\sigma_a\in I_1$. Then $A(a,\lambda_{I},\vec{\chi})=s_{1,I}(\sigma_a,\lambda_{I})$. Hence, with the use of (\ref{16})
\begin{eqnarray}\label{20}
\left \langle \rho_{I} A(a,\lambda_{I},\vec{\chi})\right\rangle_{I} = \frac{1}{2^4T\left(1-\frac{4}{T}\right)}\left( \int_{-\frac{1}{4}}^{-\frac{1}{T}} d x_{I} + \int_{\frac{1}{T}}^{\frac{1}{4}} d x_{I}\right)\prod_{k=1}^3 \int_{-1}^1 d\mu_{k,I}\times\nonumber\\
\int_{-T}^{T} d\tau_{I} \sum_{n_{I}=0}^1 \left\{n_{I}\sgn\left(\sigma_a+1-\mu_{1,I} \right)-\delta_{0,n_{I}} \right\}\sgn \left( T\Delta_T(f_{I}(x_{I}))-\tau_{I} \right)
\end{eqnarray}
From (\ref{13}) it already follows that the $\tau_{I}$ integral in (\ref{20}) equals $2T\Delta_T(f_{I}(x_{I}))$. So let us look at the $\mu$ integrals and the $n_{I}$ sum. 
Before entering into more details let us note that, lloking at (\ref{15}), because $\sigma_a \in I_1$ we have $\left(\sigma_a + 1\right) \in [-1,1]$ and so
\begin{eqnarray}\label{20a}
\int_{-1}^1 d\mu \,\sgn\left(\sigma_a + 1 - \mu\right) =\int_{-1}^{\sigma_a+1 } d\mu- \int_{\sigma_a + 1}^1 d\mu = 2\left(\sigma_a+1\right) 
\end{eqnarray}
We subsequently see, because $\int_{-1}^{+1 } d\mu_{2,I}=\int_{-1}^{+1 } d\mu_{3,I}=2$, together with $\int_{-1}^{+1 } d\mu_{1,I}=2$,
\begin{eqnarray}\label{21}
 \sum_{n_{I}=0}^1\left(\prod_{k=1}^3 \int_{-1}^1 d\mu_{k,I}\right) \left\{n_{I}\sgn\left(\sigma_a+1-\mu_{1,I} \right)-\delta_{0,n_{I}} \right\} =\nonumber\\
 2^3  \sum_{n_{I}=0}^1 \left[  n_{I}(\sigma_a+1)-\delta_{0,n_{I}}\right]=\nonumber\\
2^3[-1+(\sigma_a+1)]=2^3 \sigma_a
\end{eqnarray}
Hence, if $K_T$ is defined by 
\begin{eqnarray}\label{22}
K_T\equiv \left( \int_{-\frac{1}{4}}^{-\frac{1}{T}} d x_{I} + \int_{\frac{1}{T}}^{\frac{1}{4}} d x_{I}\right)\Delta_T(f_{I}(x_{I}))
\end{eqnarray}
then, $\left \langle \rho_{I} A(a,\lambda_{I},\vec{\chi})\right\rangle_{I}=\frac{\sigma_a K_T}{\left(1-\frac{4}{T}\right)}$ when $\sigma_a \in I_1$. 

Let us now suppose, $\sigma_a \in I_3$.  Looking at (\ref{15}) we have $\sigma_a - 1 \in [-1,1]$. 
Hence, only $\iota_3(\sigma_a)=1$ and hence, $A(a,\lambda_I,\vec{\chi})=s_{3,I}(\sigma_a,\lambda_{I})$.  The function $s_{3,I}(\sigma_a,\lambda_{I})$ is defined in (\ref{16}). This state of affairs implies,
\begin{eqnarray}\label{23}
\left \langle \rho_{I} A(a,\lambda_{I},\vec{\chi})\right\rangle_{I} = \frac{1}{2^4T(1-\frac{4}{T})}\left( \int_{-\frac{1}{4}}^{-\frac{1}{T}} d x_{I} + \int_{\frac{1}{T}}^{\frac{1}{4}} d x_{I}\right)\prod_{k=1}^3 \int_{-1}^1 d\mu_{k,I}\times\nonumber\\
\int_{-T}^{T} d\tau_{I} \sum_{n_{I}=0}^1 \left\{n_{I}\sgn\left(\sigma_a-1-\mu_{3,I} \right)+\delta_{0,n_{I}} \right\}\sgn \left( T\Delta_T(f_{I}(x_{I}))-\tau_{I} \right)
\end{eqnarray}
In the case that $\sigma_a \in I_3$, we see
\begin{eqnarray}\label{23a}
\int_{-1}^1 d\mu \,\sgn\left(\sigma_a - 1 - \mu\right) =\int_{-1}^{\sigma_a-1 } d\mu- \int_{\sigma_a - 1}^1 d\mu = 2\left(\sigma_a-1\right) 
\end{eqnarray}
Again we note, $\int_{-1}^{+1 } d\mu_{2,I}=\int_{-1}^{+1 } d\mu_{3,I}=2$, together with $\int_{-1}^{+1 } d\mu_{1,I}=2$.
\begin{eqnarray}\label{23b}
 \sum_{n_{I}=0}^1\left(\prod_{k=1}^3 \int_{-1}^1 d\mu_{k,I}\right) \left\{n_{I}\sgn\left(\sigma_a-1-\mu_{1,I} \right)+\delta_{0,n_{I}} \right\} =
\nonumber\\
 2^3  \sum_{n_{I}=0}^1 \left[  n_{I}(\sigma_a-1)+\delta_{0,n_{I}}\right]=\nonumber\\
2^3[1+(\sigma_a-1)]=2^3 \sigma_a
\end{eqnarray}
With $K_T$ given in (\ref{22}) we find 
$\left \langle \rho_{I} A(a,\lambda_{I},\vec{\chi})\right\rangle_{I}=\frac{\sigma_a K_T}{\left(1-\frac{4}{T}\right)}$. 
\newpage
Finally let us look at the case where $\sigma_a \in I_2$. Here we have $\sigma_a \in [-1,1]$. 
So, refering to (\ref{16})
\begin{eqnarray}\label{24}
\left \langle \rho_{I} A(a,\lambda_{I},\vec{\chi})\right\rangle_{I} = 
\nonumber\\
\frac{1}{2^4T(1-\frac{4}{T})}\left( \int_{-\frac{1}{4}}^{-\frac{1}{T}} d x_{I} + \int_{\frac{1}{T}}^{\frac{1}{4}} d x_{I}\right)\prod_{k=1}^3 \int_{-1}^1 d\mu_{k,I}
\int_{-T}^{T} d\tau_{I} \sum_{n_{I}=0}^1 \sgn\left(\sigma_a - \mu_{2,I}\right)
\end{eqnarray}
The result of integration in (\ref{24}) is that  $\left \langle \rho_{I} A(a,\lambda_{I},\vec{\chi})\right\rangle_{I}=\frac{\sigma_a \left(1-\frac{4}{T}\right)}{\left(1-\frac{4}{T}\right)}= \sigma_a$ and $T \sim $ sufficiently large number.

\subsubsection{The integral $K_T$}
In two cases of $\sigma_a$ we have $\left \langle \rho_{I} A(a,\lambda_{I},\vec{\chi})\right\rangle_{I}=\frac{\sigma_a K_T}{1-\frac{4}{t}}$ and $K_T$ is defined in (\ref{22}). For the ease of notation let us write $T=n$. Let us repeat the definition of the $K$ integral 

\begin{equation}
K_n=\left( \int_{-\frac{1}{4}}^{-\frac{1}{n}}dx+\int_{\frac{1}{n}}^{\frac{1}{4}} dx \right)\Delta_n\left(x^2 - (1/n^2) \right)
\end{equation}
The integral we want to discuss here is then re-written, using $\Delta_n\left(x^2 - (1/n^2) \right)$ defined in (\ref{12})  as
\begin{equation}\label{K16}
K_n=\frac{2}{\pi}\int_{-\frac{1}{4}}^{-\frac{1}{n}} \frac{dx}{1+n^2(x^2-(1/n^2))^2}+\frac{2}{\pi}\int_{\frac{1}{n}}^{\frac{1}{4}} \frac{dx}{1+n^2(x^2-(1/n^2))^2}
\end{equation}

\subsubsection{Upper $K_n$ limit}
Now let us take, $y=x^2-(1/n^2)$. The upper limit of $y$ is, $\frac{1}{4^2}-\frac{1}{n^2}$, $n>>16$,  while the lower limit is $0$. 
Hence, for negative $x$, we have $x=-\sqrt{y+\frac{1}{n^2}}$. For positive $x$, we see, $x=\sqrt{y+\frac{1}{n^2}}$. Hence, noting $dx=\pm \frac{dy/2}{\sqrt{y+\frac{1}{n^2}}}$, in terms of $y$ we can write for the two terms in $K_n$
\begin{equation}
K_n=-\frac{2}{\pi}\int_{\frac{1}{4^2}-\frac{1}{n^2}}^0\frac{dy}{1+n^2y^2}\frac{1/2}{\sqrt{y+\frac{1}{n^2}}}+\frac{2}{\pi}\int^{\frac{1}{4^2}-\frac{1}{n^2}}_0\frac{dy}{1+n^2y^2}\frac{1/2}{\sqrt{y+\frac{1}{n^2}}}
\end{equation}
Hence,
\begin{equation}\label{K17}
K_n= \left(\frac{2}{\pi}\right) 2 \int_{0}^{\frac{1}{4^2}-\frac{1}{n^2}}\frac{dy}{1+n^2y^2}\frac{1/2}{\sqrt{y+\frac{1}{n^2}}}
\end{equation}
Let us in the first place try to find the upper limit of $K_n$ from the previous equation. Note, for $n>4$ that $y+\frac{1}{n^2} \geq \frac{1}{n^2}$, hence, $\frac{1}{\sqrt{y+\frac{1}{n^2}}}\leq n$, given $\frac{1}{4^2}-\frac{1}{n^2} \geq y \geq 0$. This implies
\begin{equation}\label{K18}
K_n \leq \frac{2}{\pi} \int_{0}^{\frac{1}{4^2}-\frac{1}{n^2}} \frac{n dy}{1+n^2y^2}\leq \frac{2}{\pi} \arctan\left[ \frac{n}{4^2}-\frac{1}{n}\right]\leq 1,\hspace{0.1cm} (n \sim  large).
\end{equation}
\subsubsection{Lower $K_n$ limit existence}
Looking at (\ref{K17}) we have in the $x$ variable before transformation
\begin{eqnarray}\label{K172}
K_n = \frac{2}{\pi} \int_{x=1/n}^{x=1/4} \frac{dx}{n^2x^4-2x^2+1+\frac{1}{n^2}}
\end{eqnarray}
This form was used by an unknown referee in a previous review as the sole basis for the rejection that $K_n$ is unequal to zero, $n$ increasingly large but finite. 
In the appendix A the closed form that was employed to accomplish the $K_n$ to zero claim is presented. 
Here we state that to our minds the demonstration in the appendix A only contributes to the CHSH approving branch of the argument. 
It does not invalidate the demonstration of the "CHSH is invalid" branch.

Let us look at (\ref{K172})and select $n$ integer odd sufficiently large. Then we may have in the above integral $(1+(1/n^2))\approx 1$
Hence,
\begin{eqnarray}\label{K173}
K_n \approx \frac{2}{\pi} \int_{x=1/n}^{x=1/4} \frac{dx}{n^2x^4-2x^2+1}
\end{eqnarray}
And so, $n^2x^4-2x^2+1\leq n^2x^4+1$, with $\min\{n^2x^4-2x^2+1\,|\,1/n \leq x \leq 1/4\}=1-(1/n^2) >0$ for $n$ sufficiently large. Obviously, $n^2x^4+1\geq 0$ for $1/n \leq x \leq 1/4$. Therefore, 
\[
(n^2x^4-2x^2+1)^{-1}\geq (n^2x^4+1)^{-1}
\]
And therefore we have
\begin{eqnarray}\label{K174}
K_n \gtrapprox \frac{2}{\pi} \int_{x=1/n}^{x=1/4} \frac{dx}{n^2x^4+1}
\end{eqnarray}
Let us furthermore write $x=\sqrt{y}$.
This gives $dx=\frac{dy}{2\sqrt{y}}$.
Performing the substitution gives us.
\begin{eqnarray}\label{H175}
K_n \gtrapprox \frac{2}{\pi} \int_{x=1/n}^{x=1/4} \frac{dy}{1+n^2y^2}\left(\frac{1/2}{\sqrt{y}}\right)
\end{eqnarray}
Please observe we have, from $x\leq 1/4$ that $4\leq \frac{1}{\sqrt{y}}$. Therefore, because, \[1<2\leq \frac{1/2}{\sqrt{y}}\] from (\ref{5}) it follows
\begin{eqnarray}\label{K176}
K_n \gtrapprox \frac{2}{\pi} \int_{x=1/n}^{x=1/4} \frac{dy}{1+n^2y^2}
\end{eqnarray}
Carefully note, using $\cot(z)=\cos(z)/\sin(z)$, that $1+\cot^2(z)=1/\sin^2(z)$. The $1+n^2y^2=1/\sin^2(z)$ transformation may be interesting. 
In the evaluation we will make use of the $\arccot$ inverse.
Both $\sin$ and $\cos$ are periodic. 
Let us assume $n $ is an odd integer. This does not change the conclusion of the work.
For clarity we will denote the employed $\arccot$ with an index $n$ as $\arccot_n$.
The $\arccot_n$ projects in a proper interval with $n$ periodicity.
E.g. when in need of negative $\cot(z)$ we can employ the intervals $\frac{n}{2}\pi\leq z\leq n\pi$, where $\cos $ is negative and $\sin$ positive. Or, we may use $\frac{3n}{2}\pi\leq z\leq 2n\pi$, where $\sin$ is negative and $\cos$ is positive, as a projection of $\arccot_n$. Note, $n$ odd.

If we then write \[y=-\frac{1}{n}\cot(z)\]
the $y\geq 0$. 
When $\frac{n}{2}\pi\leq z\leq n\pi$, or, $\frac{3n}{2}\pi\leq z\leq 2n\pi$, the previous is true. It can be noted that \[dy=\frac{1}{n}\frac{1}{\sin^2(z)}dz\]  
This follows from, $ \frac{d}{dz}\cot(z)=\frac{-1}{\sin^2(z)}$.
Furthermore, we have, $1+n^2y^2=1+\cot^2(z)=\frac{1}{\sin^2(z)}$. So,
\begin{eqnarray}\label{K177}
K_n \gtrapprox \frac{2}{\pi}\frac{1}{n} \int_{x=1/n}^{x=1/4} \frac{dz}{\sin^2(z)}\frac{1}{\frac{1}{\sin^2(z)}} 
\end{eqnarray}
Or equally
\begin{eqnarray}\label{K178}
K_n \gtrapprox \frac{2}{\pi}\frac{1}{n} \int_{x=1/n}^{x=1/4} dz
=\frac{2}{\pi}\frac{1}{n}\left\{z(1/4)-z(1/n) \right\}
\end{eqnarray}
We then may note that $z$ is the $\arccot_n$ of $-ny$. So \[z=\arccot_n(-ny)=\arccot_n(-nx^2).\]
This then gives in turn that
\begin{eqnarray}\label{K179}
K_n \gtrapprox \frac{2}{\pi}\frac{1}{n}\left\{\arccot_n(-n/16)-\arccot_n(-1/n) \right\}
\end{eqnarray}
Now, if $n$ is finite large \& odd, then $\arccot_n(-n/16)\uparrow n\pi$ in the interval $\frac{n}{2}\pi\leq z\leq n\pi$. The $\cos$ is negative here and the $\sin $ approaches zero from the positive side. Similarly, $\arccot_n(-1/n)\downarrow \frac{n}{2}\pi$. 
Therefore we may write (\ref{K179}) as
\begin{eqnarray}\label{K1710}
K_n \gtrapprox \frac{2}{\pi}\frac{1}{n} \left\{\arccot_n(-n/16)-\arccot_n(-1/n) \right\}\approx \frac{2}{\pi}\frac{1}{n}\left\{n\pi-\frac{n}{2}\pi \right\}
=\\\nonumber
\frac{2}{\pi}\frac{1}{n}\frac{n}{2}\pi= 1
\end{eqnarray}
A similar story is true for  $\frac{3n}{2}\pi\leq z\leq 2n\pi$.
The conclusion from the mathematics in this section, leading to $K_n\gtrapprox 1$, is that one may try to find a lower limit, despite the closed form in appendix A that started from the same integral (\ref{K172}). 

Note please that we are looking here at the support for the branch "CHSH is not true". 
This is but one branch of the argument in concrete mathematical incompleteness. In the appendix the reader may find additional support for the claim that "CHSH is true" based on a closed form where $K_n$ is vanishing.
Note that most likely the axiom of choice, allowing the selection of the $\arccot_n$ function, is behind the $K_n\gtrapprox 1$. 
\subsubsection{Lower $K_n$ limit computation}
The lower limt in $K_n$ can be found, looking at, $1+n^2y^2\leq 1+ \epsilon^2 + n^2y^2$, hence,
\[
\frac{1}{1+n^2y^2} \geq \left( \frac{1}{1+\epsilon^2}\right) \left\{ \frac{1}{1+n^2\left( \frac{y}{\sqrt{1+\epsilon^2}}\right)^2}\right\}.
\]
Let us, in the second place, take $z=y/\sqrt{1+\epsilon^2}$, then, with $dz=dy/\sqrt{1+\epsilon^2}$ we can rewrite the lower limit like
\begin{equation}\label{K19}
K_n \geq \frac{2}{\pi} \frac{1}{\sqrt{1+\epsilon^2}}\int_0^{z_{max}}\frac{n\,dz}{1+n^2z^2}\frac{1}{\sqrt{1+n^2z\sqrt{1+\epsilon^2}}}
\end{equation}
together with, $z_{max}=\left( \frac{1}{4^2}-\frac{1}{n^2} \right)/\sqrt{1+\epsilon^2}>0$. Note, $-1\leq \frac{2}{\pi} \arctan(x) \leq 1$ for all $x \in \mathbb{R}\cup\{-\infty,\infty\}$. With $\arctan$ the inverse function of the function $-\infty \leq \tan(x) \leq \infty$, with,$-\frac{\pi}{2}\leq x \leq \frac{\pi}{2}$ is intended. Using, $\frac{d}{dz}\arctan(nz)=\frac{n}{1+n^2z^2}$ we are able to write
\begin{equation}\label{K20}
K_n \geq \frac{2}{\pi} \frac{1}{\sqrt{1+\epsilon^2}}\int_0^{z_{max}}dz \left( 
\frac{d}{dz}\arctan(nz)\right)\left[ 1+n^2z\sqrt{1+\epsilon^2} \right]^{-1/2}
\end{equation}
Because $\arctan(0)=0$ and we have $0\times n^2=0$ when $n\sim $ sufficiently large finite number, it follows that the constant factor, $C_n$ in a partial integration treatment of the right hand of (\ref{K20}) looks like
\begin{eqnarray}\label{constant}
C_n=\frac{2/\pi}{\sqrt{1+\epsilon^2}}\left\{ \arctan\left[ \left( \frac{n}{4^2}-\frac{1}{n}\right)/\sqrt{1+\epsilon^2}\right]\left[ 1+n^2z_{max}\sqrt{1+\epsilon^2} \right]^{-1/2}\right.
\nonumber\\
-\left.\arctan(0)[1+(n^2 \times 0)]^{-1/2} \right\}
\end{eqnarray}
and $1+n^2z_{max}\sqrt{1+\epsilon^2} = 1+\frac{n^2}{4^2}-1$. Hence, $\left[ 1+n^2z_{max}\sqrt{1+\epsilon^2} \right]^{-1/2}=\frac{1}{\sqrt{\frac{n^2}{4^2}}}=\frac{4}{n}$. 
This implies, the constant factor 
\[C_n=\frac{2/\pi}{\sqrt{1+\epsilon^2}}\arctan\left[ \left( \frac{n}{4^2}-\frac{1}{n}\right)/\sqrt{1+\epsilon^2}\right]\frac{4}{n} \approx 0^{+}  \] for, $n\sim$ sufficiently large number. 

\begin{figure}
\center
\caption{{\it Plot of the integrand of the integral in (\ref{K20a}). Parameters $(n,h)=(3.5\times 10^{17},5.3\times 10^{-6})$ are given in the program parameter statements. We find $1\geq K_n\gtrapprox 0.9671$. This result is computed with the program to be found in appendix B.}}
\includegraphics[scale=0.55]{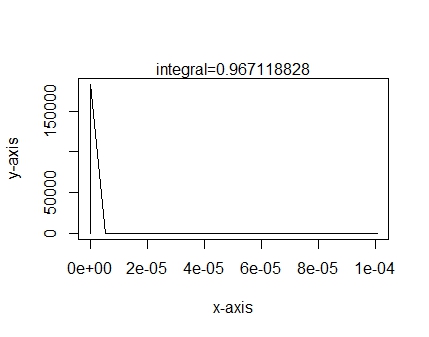}
\label{fig1}
\end{figure}

So, under $n \sim$ sufficiently large number, for $z\neq 0$, we see from $-1\leq \frac{2}{\pi} \arctan(nz) \leq 1$ that the extremes $-1$ and $+1$ are quickly approximated. 
In turn, the partial integration of the right hand of (\ref{K20}), finally looks like
\begin{equation}\label{K20a}
K_n \geq C_n -\frac{1}{\sqrt{1+\epsilon^2}}\frac{2}{\pi} \int_0^{z_{max}}dz\, \arctan(nz)\left( \frac{d}{dz}\left[ 1+n^2z\,\sqrt{1+\epsilon^2} \right]^{-1/2} \right)
\end{equation}
Hence,  when $z > 0$ from $-\frac{2}{\pi} \arctan(nz) \approx -1$ and $C_n \approx 0^{+}$, for $n \sim$ large,
\begin{equation}\label{K21}
K_n \gtrapprox -\frac{1}{\sqrt{1+\epsilon^2}} \int_0^{z_{max}}dz \left( \frac{d}{dz}\left[ 1+n^2z\,\sqrt{1+\epsilon^2} \right]^{-1/2} \right)
\end{equation}
Note that the step from (\ref{K20a}) to (\ref{K21}) is supported by the following subsubsections plus a result of numerical study represented in figure-\ref{fig1}. In the fortran code that can be obtained as a separate file, the first initial statements are there to search for a way to catch, so to speak, the singularity. The way it is done is given in the computer code.
We believe that the computer code represents an essential step in the demonstration and therefore its reference is included in the paper. 
Below the evidence is displayed in a figure\footnote{The reader can find the computer program behind the figure in appendix B}.
\newpage

\subsubsection{Mean value theorem}
A way to look at the possibility $1\geq K_n >1-\epsilon$ is to approximate $K_n$ with the use of the first mean value theorem for definite integrals. 
If $f(x)$ and $g(x)$ are $\mathbb{R}^{+}$ functions in interval $(a,b)$ and we look at an interval $(a,b)$, i.e. $f:(a,b)\rightarrow \mathbb{R}^{+}$ and $g:(a,b)\rightarrow \mathbb{R}^{+}$ then there is a, $c \in (a,b)$ such that
\[
\int_a^b f(x)g(x) dx =f(c)\int_a^b g(x) dx
\]
Furthermore we write the right hand side of (\ref{K20a}) such that $K_n \geq C_n + L_n$. Here $C_n$ is defined as in (\ref{constant}), while 
\begin{equation}\label{ad1}
L_n = -\frac{1}{\sqrt{1+\epsilon^2}}\frac{2}{\pi} \int_0^{z_{max}}dz\, \arctan(nz)\left( \frac{d}{dz}\left[ 1+n^2z\,\sqrt{1+\epsilon^2} \right]^{-1/2} \right)
\end{equation}
Then obviously we may derive
\begin{equation}\label{ad2}
L_n=\frac{1}{\sqrt{1+\epsilon^2}}\int_0^{z_{max}}dz f(z)g(z)
\end{equation}
where, $f(z)=(2/\pi)\arctan(nz)$, continuous and positive in $(0,z_{max})$ and 
\[
\forall_{z \in (0,z_{max})}\,\,
g(z)=\frac{1}{2}\frac{n^2\sqrt{1+\epsilon^2}}{\left(1+n^2 z \sqrt{1+\epsilon^2}\right)^{3/2}}=-\frac{d}{dz}\left[ 1+n^2z\,\sqrt{1+\epsilon^2} \right]^{-1/2} \geq 0
\]
continuous and integrable (finite integral) in $(0,z_{max})$, with, 
\[z_{max}=\left( \frac{1}{4^2}-\frac{1}{n^2} \right)/\sqrt{1+\epsilon^2}>0.\] We have that $\forall_{z\in (0,z_{max})}f(z)g(z)\geq 0$. So, we may apply the first mean value theorem for definite integration (n large but finite). This means that there is a $\zeta \in (0,z_{max})$ such that 
\begin{equation}\label{ad3}
\frac{1}{\sqrt{1+\epsilon^2}}\int_0^{z_{max}}dz f(z)g(z)=f(\zeta)\frac{1}{\sqrt{1+\epsilon^2}}\int_0^{z_{max}}dz g(z)
\end{equation}
Hence, for sufficient large but finite $n$, we see $f(\zeta)=(2/\pi)\arctan(n\zeta) \approx 1$, such that 
\begin{equation}\label{ad4}
L_n \approx-\frac{1}{\sqrt{1+\epsilon^2}} \int_0^{z_{max}}dz  \frac{d}{dz}\left[ 1+n^2z\,\sqrt{1+\epsilon^2} \right]^{-1/2} =\frac{1-(4/n)}{\sqrt{1+\epsilon^2}}
\end{equation}
Hence, $ 1 \geq K_n \geq  C_n + L_n \gtrapprox C_n +\frac{1-(4/n)}{\sqrt{1+\epsilon^2}}$ gives what is described in the next paragraph. 

It is noted that the first mean value theorem for definite integration is based on the intermediate value theorem. 
In Bishops constructive analysis, there is serious doubt about the intermediate value theorem \cite[ introduction section]{BISH}. Modern developments show that a weaker version of the theorem can be maintained but without axiom of choice \cite{Hen}. 
\subsubsection{Wrapping it up}\label{PartD}
Returning to equation (\ref{K21}). This then gives, using $z_{max}=\left( \frac{1}{4^2}-\frac{1}{n^2} \right)/\sqrt{1+\epsilon^2}$,
\begin{equation}\label{K22}
1 \geq K_n \gtrapprox -\frac{1}{\sqrt{1+\epsilon^2}}\left[\frac{4}{n} -1  \right]  \rightarrow  \frac{1}{\sqrt{1+\epsilon^2}},
\end{equation}
under the condition, $n \sim$ large number. 
Hence, we may conclude that: \[1 \geq  K_{n} \gtrapprox \frac{1}{\sqrt{1+\epsilon^2}}.\]
for $n$ large. This leads us to,  $K_n \approx 1$, where $\epsilon^2$ can be arbitrary small positive real and $n$ sufficently large. 

\section{Result}
Returning to $\sigma_a \in I_1$ and $\sigma_a \in I_3$, it is found that approximately we may write $\left \langle \rho_{I} A(a,\lambda_{I},\vec{\chi})\right\rangle_{I}\approx \sigma_a$, because under $T=n$ sufficiently large, $K_T \approx 1$. 
Moreover under $T \sim $ sufficiently large number we also see that for $\sigma_a \in I_2$ that $\left \langle \rho_{I} A(a,\lambda_{I},\vec{\chi})\right\rangle_{I}\approx \sigma_a$. Hence, because a similar evaluation for $B$ can take place
\begin{eqnarray}\label{R1}
\left \langle \rho_{I} A(a,\lambda_{I},\vec{\chi})\right\rangle_{I}\approx \sigma_a \nonumber \\
\left \langle \rho_{II} B(b,\lambda_{II},\vec{\chi})\right\rangle_{II}\approx \sigma_b
\end{eqnarray}
Because using (\ref{18}) and our previous result, we are allowed to write 
\begin{eqnarray}\label{R2}
E(a,b)=\left \langle \left \langle \rho_{I} A(a,\lambda_{I},\vec{\chi})\right\rangle_{I}\left \langle \rho_{II}B(b,\lambda_{II},\vec{\chi}) \right\rangle_{II}\rho_{Norm} \right\rangle_{\mathbb{R}^3}\approx \nonumber \\
\left \langle \sigma_a (\vec{\chi}) \sigma_b(\vec{\chi}) \rho_{Norm} \right\rangle_{\mathbb{R}^3} 
\end{eqnarray}
This implies, together with (\ref{8a})  that 
\begin{eqnarray}\label{R3}
E(a,b)\approx \sum_{i=1}^3 \sum_{j=1}^3 a_i b_j \delta_{i,j} = \sum_{j=1}^3 a_j b_j
\end{eqnarray}
The latter equation concludes the refutation part of the present paper.

\section{Conclusion}
In our paper, under locality \cite{2}, \cite{1}, we have construed a model that violates the CHSH but must, by design, not be able {\it in any way} to violate the $S(a,b,c,d)\leq 2$. 
We note that the local hidden variables physical picture is that variables with the index $\alpha=I$ reside in measurement instrument A and $\alpha=II$ reside in measurement instrument B. 
This is perfectly in agreement with a possible physics behind the correlation. 
The $\chi$ Gaussian variables can be seen as being carried by the particles to the respective measurement systems. This makes a perfectly valid physical possibility. 
Note however, that the physics is unimportant. 
We claim there is inconsistency in mathematics. 
We note, in addition that \cite{12} demonstrates that a local algorithm computational violation is possible as well.

In the paper it was derived that a model with $S(a,b,c,d) > 2$ can be obtained observing all conditions for a local model. I.e.
$
E(a,b)\approx \sum_{j=1}^3 a_j b_j.
$
was derived using {\it local} modeling. 
The approximation can be arbitrary close.
As can be easily checked, our result is unrelated to a quantum mechanical violation of the inequality. 

The demonstration is based on valid mathematical operations like partial integration and the mean value theorem for definite integrals. 
If there are reasons to exclude those from the repertoire of operations then we may definitely question the claimed generality of the Bell theorem. 

The numerical computations presented in the appendix support the fact that the integral under study is unequal to zero.
To us, a no-go for our derivations and computations means: delivering proof of an  error in the mathematics and numerics we employed.

In accordance with \cite{FriedLast} we think we have demonstrated that Bell's theorem is unprovable i.e. negation incomplete. 
This means incompleteness from the standard rulebook of mathematics \cite{FriedLect}. 
Without further disproof, Bells "theorem" is unprovable in the sense of G{\" o}del's phenomenon in concrete mathematics. No conclucions about the physics of entanglement can be obtained from it.


\newpage

\appendix{\underline{Appendix A:}}
The claim that $K_n$ goes to zero for large $n$ was presented to the authors in a previous review process. The claim of the unknown referee was that the integral $K_n$ had a closed form solution, as presented below. 
The reader can in this appendix verify the CHSH side of the claim. In the main text we have demonstrated $K_n\gtrapprox 1$. Here it is rejected.
We start again with
\begin{eqnarray}\label{1}
K_n = \frac{2}{\pi n^2} \int_{x=1/n}^{x=1/4}\frac{dx}{x^4-2\frac{x^2}{n^2}+\frac{n^2 +1}{n^4}} = \frac{2}{\pi} \int_{x=1/n}^{x=1/4} \frac{dx}{n^2x^4-2x^2+1+\frac{1}{n^2}}
\end{eqnarray}
The closed form is based on the subsequent function, $F_n$
\begin{eqnarray}\label{1a}
F_n(x)=\frac{n\sqrt{2}}{\pi(n^2+\sqrt{1+n^2} +1)}(G_n(x)+H_n(x))
\end{eqnarray}
with
\begin{eqnarray}\label{1b}
G_n(x)=n \log\left( \frac{U_{n,+}(x) +\sqrt{1+n^2}+1}{U_{n,-}(x)+\sqrt{1+n^2}+1}\right)
\end{eqnarray}
with
\[
U_{n,\pm}(x)=
{n^2(x^2(\sqrt{1+n^2} +1)+1)\pm n\sqrt{2}\, x \left(\sqrt{1+n^2}+1\right)^{(3/2)}}
\]
together with
\begin{eqnarray}\label{1c}
H_n(x)=2(\sqrt{1+n^2}+1)(\arctan(I_n(x))-\arctan(J_n(x)))
\end{eqnarray}
The $I_n(x)$ and $J_n(x)$ are defined as
\begin{eqnarray}\label{1d}
I_n(x)=\frac
{\sqrt{\sqrt{1+n^2}+1}+nx\sqrt{2}}{\sqrt{\sqrt{1+n^2}-1}}\\\nonumber
J_n(x)=\frac
{\sqrt{\sqrt{1+n^2}+1}-nx\sqrt{2}}{\sqrt{\sqrt{1+n^2}-1}}
\end{eqnarray}
This set of definitions (\ref{1a})-(\ref{1d}) define the $F_n(x)$ and we have 
\begin{eqnarray}\label{1u}
K_n=F_n(1/4)-F_n(1/n)
\end{eqnarray}
The reader can check that indeed the $K_n$ expressed in (\ref{1u}) approaches zero for increasing $n$. However, how to explain the contrast with $K_n\gtrapprox 1$ such as given in the main text.
We believe that this is the ultimate example of G{\"o}del negation incompleteness in concrete mathematics. 
\newpage
\appendix{\underline{Appendix B:}} 
This result can be found in figure-\ref{fig1} above
With the particular parameters in the code we get $1\geq K_n \gtrapprox 0.9671$.
\begin{verbatim}
      program testAtAr
      integer nmax, n, m,j,k
      real*8 h,xx,funfArr,nlim,f2,ffHelp
      real*8 pw
      real*8 eps,beta,fact,y0,ystart,yfin
      real*8 g1,g2,f,f0,resl, integral
      parameter(nlim=3.5e17)
      parameter(nmax=50)
      parameter (m=100)
      parameter (h=5.3e-6)
      real*8 ffArray(m),xxVar(m)
c output for plot
      open(1,
     +file='res.txt'
     +,status='unknown')
      open(2,
     +file='xres.txt'
     +,status='unknown')
      open(3,
     +file='yres.txt'
     +,status='unknown')
      write(1,*)0.0
      write(2,*)0.0
      eps=1/nlim
      f0=(nlim*nlim)*dsqrt(1.0+(eps*eps))
c determine the proper starting point given the integration
c interval h to 'catch' the singularity at a given n
      beta=-(2.6)/(3.0)
      y0=1.0/dsqrt(1.0+(eps*eps))
      pw=-(2.0)/(3.0)
      f2=(2.0)**pw
      y0=y0*(f2*(nlim**beta)-(1/(nlim*nlim)))
      ystart=y0-(9.0*h)
      yfin=y0+(20.0*h)
c the yfin is there to not waste too much iterations 
      xx=ystart
      j=0
  10  continue
              if(xx.gt.0) then
                j=j+1
                xxVar(j)=xx
                ffHelp=funfArr(xx,nlim)
                g1=ffHelp/dsqrt(1.0+(eps*eps))
                f=f0*xx
                g2=f0/((f+1.0)**(1.5))
                ffArray(j)=(g1*g2)/2.0
                write(1,*)ffArray(j)
                write(2,*)xxVar(j)
              endif
              xx=xx+h
      if (xx.lt.yfin) go to 10
      write(*,*) 'number of iterations=',j
c integration  
      resl = integral(ffArray,h,j)
      write(*,*)'integral=',resl
      write(3,*) resl
      close(unit=3)
      close(unit=2)
      close(unit=1)
      stop
      end
      real*8 function integral(ffArray,h,n)
      integer i,j,k,n,m
      parameter (m=100)
      real*8 sum,h,ffArray(m)
      sum=0
      do 10 j=1,n
  10       sum=sum+(ffArray(j)*h)
      integral=sum
      return
      end
      real*8 function funfArr(xx,nlim)
      real*8 xx,nlim
      real*8 pi,y,z
       z=nlim*xx
       pi=4*atan(1.0)
       y=(2/pi)*atan(z)
       funfArr=y
      return
      end
\end{verbatim}
\end{document}